\documentclass[12pt]{article}
\usepackage{authblk}
\usepackage[top=0.6in,bottom=0.4in,left=0.7in,right=0.2in]{geometry}
\usepackage{setspace}
\usepackage{multirow}
\usepackage{amssymb}
\usepackage{graphicx}
\usepackage{hyperref}
\usepackage{caption}

\hypersetup{
	colorlinks=true,
	linkcolor=blue,
	citecolor=blue,
	filecolor=magenta,
	urlcolor=cyan,
}
\begin{document}
	\title{Emergence of spin-phonon coupling in a Gd-doped Y$_2$CoMnO$_6$ double perovskite oxide: a combined experimental and ab-initio study}
	
	\author[1]{Anasua Khan}
 \author[2]{Debdatta Banerjee}
    \author[3]{Divya Rawat}
	\author[1]{T.K Nath}
    \author[3]{Ajay Soni}
     \author[2]{Swastika Chatterjee}
	\author[1]{A.Taraphder}

	\affil[1]{Department of Physics, Indian Institute of Technology, Kharagpur-721302, West Bengal, India}
	\affil[2]{National Centre for High-Pressure Studies and Department of Earth Sciences, Indian Institute of Science Education and Research Kolkata, Nadia- 741246, West Bengal, India}
    \affil[3]{School of Physical Sciences, Indian Institute of Technology Mandi, Mandi-175005, Himachal Pradesh, India}

	\setcounter{Maxaffil}{0}
	\renewcommand\Affilfont{\itshape\small}
	\date{}

	\maketitle
	\begin{abstract}
		
 We present Raman spectroscopy results backed by first-principles calculations and investigate the nature of possible spin-phonon coupling (SPC) in a Gd-doped Y$_2$CoMnO$_6$ (YGCMO) double perovskite oxide. The influence of Gd substitution, A-site ordering, and anti-site disorder is also studied. YGCMO exhibits anti-site disorder leading to both ferromagnetic (between Co and Mn) and antiferromagnetic interactions (Co-Co, Mn-Mn, Gd-Co/Mn), making the SPC quite intriguing. An analysis of the temperature-dependent phonon frequencies for the stretching modes of YGCMO indicates that SPC here possibly emerges from the simultaneous presence of competing ferromagnetic and antiferromagnetic interactions. The SPC strength comes out to be 0.29 cm$^{-1}$. Our density functional theory (DFT) calculations show that Phonon modes shifted towards lower frequency with Gd doping. Similarly, A-site ordring and anti-site disorder significantly alter the Raman spectra. Experimental findings are also corroborated by first-principles DFT calculations, which indicate that anti-site disorder and Gd doping enhances SPC in YGCMO. This implies a strong influence of A-site cationic radii, and B-site (Co/Mn) ordering on SPC in the bulk double perovskite systems. The phonon dynamics of YGCMO are, therefore, correlated with magnetic ordering, indicating potential applications in spintronics devices. 
\end{abstract}
	
\section{Introduction}
	
Multifunctional materials with correlated electronic structures have of late witnessed a surge in interest due to their inherent fundamental physics \cite{int_1} (like frustrated magnetism and various spin states) and potential devices application (such as memories \cite{int_2} and spintronics \cite{int_3}). The strong correlation between the spin, charge, orbital, and lattice degrees of freedom is responsible for such emergent properties \cite{int_4, int_5}. In this regard, the compounds with the generic formula R$_2$BMnO$_6$ (R= rare earth B = transition metal) have been found to hold great promise.

In general, the ordered double perovskite (DP) R$_2$BMnO$_6$ materials are found to be ferromagnetic (FM) in nature by virtue of the B$^{2+}$-O-Mn$^{4+}$ superexchange interaction \cite{int_6, int_7}. Previous experimental and theoretical research on this class of materials reports the emergence of several novel functionalities, namely, magnetocapacitance, magnetoresistance, relaxor ferroelectricity, and multiferroicity \cite{int_9,int_10,int_11,int_12}. Besides, the nature of order/disorder at the B and Mn sites in these DP materials has been found to greatly influence the electronic, magnetic, vibrational, and dielectric properties \cite{int_13,int_14,int_15,int_5,int_17}. Anti-site disorder (ASD) may also result in the suppression of ferromagnetism in some materials. This is because ASD induces B$^{2+}$-O-B$^{2+}$ and Mn$^{4+}$-O-Mn$^{4+}$ super-exchange interactions, which result in short-range anti-ferromagnetic (AFM) interactions \cite{EXP_1}. 

 Recently, the DPs containing smaller rare earth cations have been of great research interest as many theoretical, as well as experimental studies report the emergence of novel multiferroic properties \cite{int_21} in these materials. Of particular interest is the Y-based double perovskite such as Y$_2$CoMnO$_6$ (YCMO) and Y$_2$NiMnO$_6$, which have been found to exhibit polar nature in their magnetic ground state \cite{int_22}. Moreover, first-principles calculations have shown that YCMO double perovskite oxides develop E* type magnetic ordering which is the source of ferroelectricity in these materials \cite{int_23, int_24}. It also exhibits some interesting properties like disorder-induced exchange bias and other multiferroic properties \cite{EXP_1}.
 
 In a strongly correlated system, the richness of the novel functionalities depends on the coupling of magnetic spins, the crystal lattice, and lattice vibrations. The interaction is complex but rich in the case of DP materials as it has two different magnetic species. So, a qualitative model based on spin-phonon coupling was proposed earlier for YCMO \cite{int_25} in order to explain magnetically induced ferroelectricity in this material. Moreover, SPC provides a fundamental background to understand various phenomena, such as magneto-thermal transport, magnetoelectric coupling, thermal Hall Effect, relaxation time in spintronics, etc. \cite{int_26, int_D}. Spin-phonon coupling also facilitates the design of different spintronic devices with low-power and high-speed operations \cite{int_S}. For instance, external magnetic fields are used to regulate magnetic refrigeration, isothermal entropy, and adiabatic temperature \cite{int_S}. Furthermore, SPC also provides a chance to modulate the functionality of transition metal oxides (TMO) \cite{int_27}. It is now well-understood that the coupling of phonons with other degrees of freedom may play a vital role in dictating several properties of these systems. We see that several studies have delved into the electronic, magnetic, optical, and transport properties of DP systems, but very few phonon-related phenomena and their interactions with other degrees of freedom have been investigated in their bulk form. It has been reported that a large spin-phonon coupling is observed in thin films of A$_2$CoMnO$_6$ (A = rare earth) double perovskite (DP) compounds, evidenced by significant softening of phonon modes below the Curie temperature (T$_C$), which is absent in their bulk counterparts \cite{int_16}. Similarly, Raman spectroscopy on La$_2$CoMnO$_6$ revealed subtle shifts in phonon frequency but substantial changes in linewidth across T$_C$, attributed to lattice effects \cite{int_viswanathan}. Kumar \textit{et al.} further investigated that the softening of the phonon mode decreases with decreasing ionic radius of A-site cation in bulk DP compounds \cite{int_16}. Conversely, Macedo Filho et al. found that in RE$_2$NiMnO$_6$ (RE = Pr, Nd, Gd, Tb, Y) compounds, the strength of spin-phonon coupling is largely independent of the A-site cation's ionic radius and is primarily governed by B-site cationic ordering \cite{int_Macedo}. Hence, investigating the nature of the “coupling" is crucial for a better understanding of these systems. 

In our previous work \cite{EXP_1}, we reported the experimental and theoretical investigation of the structural, electronic, and magnetic properties of 50\% Gd-substituted YCMO. 50\% doping at the rare-earth site helps to maintain a balanced charge state and structural stability of the doped compound. The substitution of larger Gd ions at A-site induces structural distortion, which might be coupled with the spin state of YGCMO. So, in this study, we explore the spin-phonon coupling (SPC) in YGCMO and the roles of Gd substitution and anti-site disorder on it using temperature-dependent Raman spectroscopy (serves as a powerful tool to investigate ordering, magnetism, and spin-phonon interactions) \cite{int_29} and first-principles density functional theory (DFT) calculations. This approach enables us to elucidate the origin of SPC, distinguishing between direct coupling and magnetostriction effects.

\section{Experimental and Computational Details}
The polycrystalline, single-phase double perovskite sample of Y$_{2-x}$Gd$_x$CoMnO$_6$ ($x$=1.0) (YGCMO) was synthesized by the well-known chemical sol-gel method. A detailed description of sample preparation can be found in Ref. \cite{EXP_1}

Raman scattering measurements were performed using Horriba Jobin Vyon LabRAM HR Evolution Raman spectrometer equipped with Czerny-turner grating (1800 gr/mm), 633 nm laser excitation of 1.25 mW power, 60s acquisition time, and Peltier cooled CCD detector. Low temperature-dependent Raman measurements were performed using a Montana cryostat in the temperature range of 4-300 K with the temperature step 20 K. An ultra-low frequency filter was used to access low-frequency Raman modes. All the spectra were fitted by the Lorentzian function to evaluate the phonon frequency of the Raman modes.

All theoretical calculations have been performed using first principles density functional theory (DFT) \cite{EXP_2, EXP_3} using projector augmented wave method (PAW) \cite{EXP_4, EXP_5} as implemented in the VASP code \cite{EXP_6, EXP_7}. The Perdew-Burke-Ernzerhof (PBE) \cite{EXP_8} formalism of the generalized gradient approximation (GGA) has been used as the exchange correlation functional in our calculations. An energy cut-off of 550 eV and kpoint mesh of 6$\times$6$\times$6 was applied using the Monkhorst-Pack method \cite{EXP_9}. The energy convergence with respect to all computational parameters was ascertained. A complete optimization of the crystal structure was performed till the forces were less than 0.001 eV/\AA. In order to take into account, the missing correlation beyond GGA, a supplemented GGA+U calculation was performed with the U-J (U: on-site Coulomb repulsion; J: Hunds coupling strength) \cite{EXP_10} value as 3 eV for Co-d, 3 eV for Mn-d \cite{int_13} and 6 eV for Gd-f-orbitals \cite{EXP_1}. Phonon frequencies and eigen vectors were determined from the dynamical matrices which were in turn determined from the force constants obtained using density functional perturbation theory (DFPT) \cite{EXP_11} as implemented in VASP and post-processed using PHONOPY \cite{EXP_12}. 

\section{Results and discussion}
\begin{figure*}[htp]
	\centering
        \includegraphics[scale=0.75]{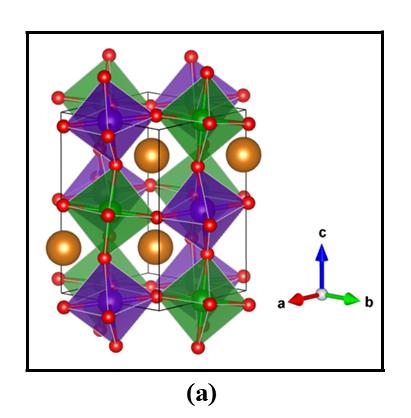}
	\includegraphics[scale=0.42]{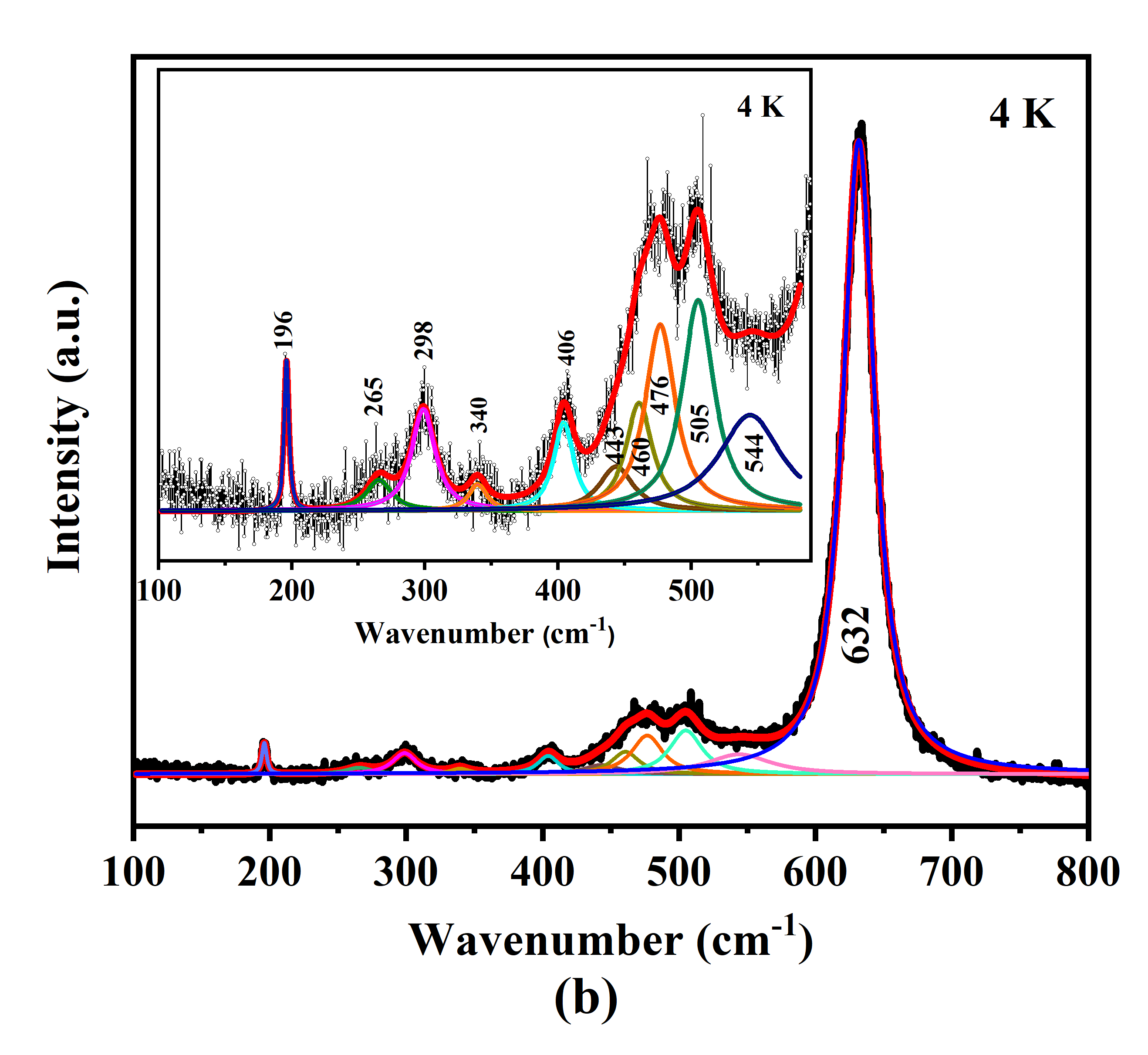}
	\caption{(a) Schematic structure of ordered YGCMO. The green and purple balls represent Co and Mn atoms, respectively. Here big golden ball represents Y/Gd atoms, sitting in the cage formed by CoO$_6$ and MnO$_6$ octahedra. (b) Raman spectra of Gd-doped YCMO at 4 K. The inset shows low wavenumber region. The red solid line shows the net fitted spectrum and blue, green, orange, cyan, and brown lines correspond to individual peaks.}
 \label{raman1}
\end{figure*}

 The detailed study on structure, oxidation states, and the magnetic states of the YGCMO compound has been reported earlier \cite{EXP_1}. The crystal structure belongs to the $P2_1/n$ space group (refer to Fig.S1 within the SI). In this symmetry, 4\textit{e} sites are generally occupied by the rare earth ions (Y/Gd) and oxygen atoms (O), whereas Co and Mn ions prefer 2\textit{c} and 2\textit{b} sites, respectively. The schematic of the structure is shown in Fig.~\ref{raman1} (a). The XPS study has shown that transition metal ions Co and Mn are present in their mixed valence states (refer to Fig.S2 within the SI), i.e., Co$^{2+}$, Co$^{3+}$, Mn$^{3+}$, and Mn$^{4+}$, respectively. The compound is found to be partially disordered with Co-Mn ions, and the calculated value of disorder becomes 38\%. Finally, the magnetic measurement reveals that the compound exhibits three different magnetic transitions: i) Transition metal ions (Co/Mn) mediated FM transition triggers at T$_C$= 95.5 K. ii) Antiferromagnetic (AFM) transition occurs at T$_N$= 47 K due to $3d-4f$ exchange interaction of Co/Mn and Gd sublattices and ASD disorder present in the sample. iii) At T$\leq$20 K, Gd$^{3+}$-O-Gd$^{3+}$ interactions become active and give rise to AFM interactions. More details are found in the supplementary information (refer to Fig.S3).

\begin{figure*}[htp]
	\centering
	\includegraphics[scale=.5]{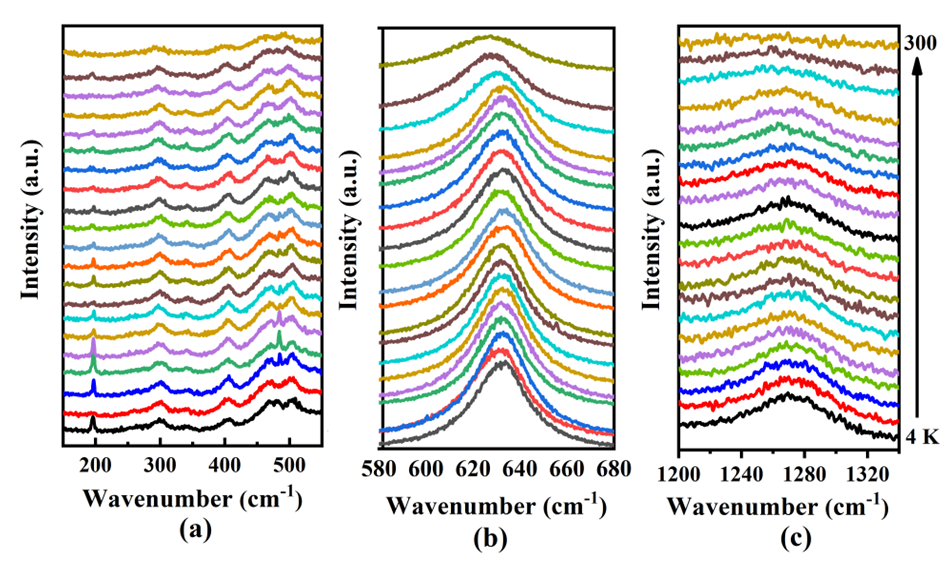}  
	\caption{Temperature-dependent Raman spectra in the (a) low wavenumber region, (b) main vibrational mode, and (c) high wavenumber region.}
 \label{raman2}
\end{figure*}

The Raman spectrum of 50\% Gd-substituted YCMO recorded at \textit{T}=4 K within the spectral range between 100 and 800 cm$^{-1}$ is shown in Fig.~\ref{raman1}(b). The absence of the most intense Raman peak of Y$_2$O$_3$ near 377 cm$^{-1}$ \cite{RD_1, RD_2} discards the presence of a secondary phase (RE$_2$O$_3$) \cite{RD_3} in the system. Generally, the low symmetry structures with space group $P2_1/n$ of double perovskites result from a distorted cubic Fm$\bar{3}$m aristotype lattice \cite{RD_4}. The group theory analysis predicted 60 first-order gamma-point phonon modes, of which 24 correspond to Raman active modes and 36 are infrared active modes. Its distribution according to the irreducible representation of 2/m factor point group is given as follows: $6T(3A_g+3B_g)+6L(3A_g+3B_g)+2\nu_1(A_g+B_g)+4\nu_2(2A_g+2B_g)+6\nu_5(3A_g+3B_g)$, which can be written as 12A$_g$+12B$_g$ \cite{int_29}, where lattice modes ($T$= translational and $L$= librational) and internal modes corresponding to oxygen octahedron ($\nu_1, \nu_2,$ and $\nu_5$) are observed below and above 380 cm$^{-1}$ \cite{int_29}. We observed eleven first-order Raman modes below 800 cm$^{-1}$ and their frequencies are determined by deconvoluting the spectrum with the Lorentzian line profile. A significant number of deconvoluted modes have been observed due to more structural distortion, primarily caused by the substitution of Gd$^{3+}$ (r = 1.22 \AA) cation at Y$^{3+}$ (r = 1.20 \AA) site. The mode noticed at higher wave number 1270 cm$^{-1}$ corresponds to the second-order overtone of breathing mode shown in Fig.~\ref{raman2} (c) \cite{RD_8}. Later, we subsequently explore the influence of Gd doping and Gd ordering on the Raman spectrum, alongside the impact of antisite disorder (ASD) on the Raman spectrum, through theoretical studies.  
                                      
In the past, Iliev \textit{et al}. \cite{int_29} have identified the Raman modes for Mn-based DP system from the semiempirical calculation, and it is in good agreement with their experimental results. Therefore, based on some of the previous works \cite{int_29, RD_7}, Raman modes at 632 cm$^{-1}$, 505 cm$^{-1}$, 340 cm$^{-1}$, 298 cm$^{-1}$, 265 cm$^{-1}$, and 196 cm$^{-1}$ observed in our experiment can be assigned to A$_g$ symmetry and also known as stretching mode (S). Here Co/Mn-O bond involves both in-plane and out-of-plane vibrations and those detected at 544 cm$^{-1}$, 476 cm$^{-1}$, and 406 cm$^{-1}$ have B$_g$ symmetry and are also known as anti-stretching mode (AS). Here O$^{2-}$ ions vibrate perpendicularly to the Co/Mn-O bond.
Generally, vibrational modes correspond to oxygen atoms connected with higher valence ions \cite{RD_5}. 
 
 Now, we focus on some first-order Raman modes and the evolution of their Raman shifts with temperature. 
The recorded temperature dependence of the Raman spectrum between 4 K and 300 K is shown in Fig.~\ref{raman2} (a), (b), and (c). According to, Granado et al.\cite{ref}, the change in phonon frequency with temperature in magnetic materials can be expressed as follows: $\omega(T)=\omega_0+\Delta\omega_{latt}+\Delta\omega_{anh}+\Delta\omega_{SPC}+\Delta\omega_{EPC}$, where $\omega_0$ is the phonon frequency at 0 K. $\Delta\omega_{latt}$ corresponds to the change in phonon wave-number due to the thermal expansion of lattice. The change in phonon frequency due to intrinsic anharmonic vibrational potential at constant volume is given by $\Delta\omega_{anh}$. $\Delta\omega_{SPC}$ and $\Delta\omega_{EPC}$ are the spin-phonon and electron-phonon coupling terms, respectively.  

$\Delta\omega_{latt}$ can be determined by the change in unit-cell volume with temperature. However, no spectral changes are observed throughout the temperature variation shown in Fig.~\ref{raman2} (a), (b), and (c), confirming that the YGCMO remains in a monoclinic structure with space group $P2_1/n$. Therefore, the observed phonon frequency shift is probably not caused by the lattice expansion with temperature. 

The anharmonicity in lattice vibration can modify the Raman shift ($\Delta\omega_{anh}$) and FWHM of the phonon modes at higher temperature. In general, the temperature dependence of phonon frequency due to anharmonic contribution is proposed by Balkanski \cite{RD_10} can be written as follows,
 \begin{equation}
	\omega (T) = \omega (0) + C \left(1 + \frac{2}{e^{\frac{\hbar\omega_0}{KT}}-1}\right)
 \label{ramaneq1}
\end{equation}  
\begin{figure*}[htp]
	\centering
	\includegraphics[scale=.5]{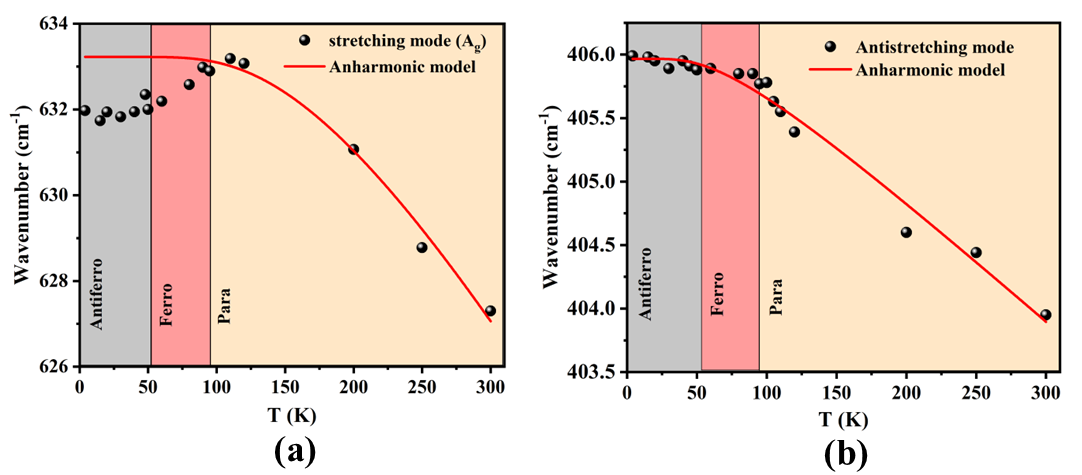}  
	\caption{Temperature-dependence of phonon positions of (a) A$_g$ stretching mode, (b) B$_g$ anti-stretching mode for YGCMO (black sphere). The blue line represents the approximation from the above anharmonic model. The error bars are the standard deviation of the peak positions as obtained from the fitting procedure.}
 \label{raman3}
\end{figure*}
where $C$ (anharmonic constant) and $\omega_0$ (position of the Raman mode at the temperature T=0 K) are the adjustable parameters. In our study, the temperature dependence of phonon frequency for two of the most representative Raman modes is analyzed using Eq.~\ref{ramaneq1} and presented in Fig.~\ref{raman3} (a-b). In the absence of any phase transition, temperature-dependent Raman mode should follow the above anharmonic relation. It is seen that the most intense Raman mode, also known as stretching mode (observed at 632 cm$^{-1}$), fits nicely with the anharmonic model above $T_C$ i.e, the compound is in the paramagnetic region, shown in Fig.~\ref{raman3}(a). This mode shows a deviation from the above model for T$\leq$T$_C$ with two distinct discontinuities at T$_C$ and T$_N$, respectively. The deviation is more pronounced at lower temperature, which implies that the phonon frequency is affected by many other factors, apart from the anharmonicity. This anomalous behavior could arise due to the renormalization of phonons induced by the ferromagnetic ordering of Co-Mn ions and results in the coupling between spin and lattice (phonon) degrees of freedom \cite{RD_11}. A similar feature is observed in several double perovskite manganite systems \cite{RD_12, RD_13}. On the other hand, the AS Raman mode observed at 406 cm$^{-1}$, fits well with the above anharmonic model throughout the investigated temperature range except near T$_C$. At lower temperatures, the phonon frequencies become nearly constant for this mode. This temperature-independent behavior of phonon frequencies at lower temperatures could happen due to a change in magnetic spin structure or weak ferromagnetic ordering \cite{RD_14}. Our previous study shows that below T$_{Gd}$ (Gd ordering temperature), Gd spins are at \textit{ab} plane, which was in any other direction \cite{EXP_1}. 
\begin{figure*}[htp]
	\centering
       \includegraphics[scale=.3]{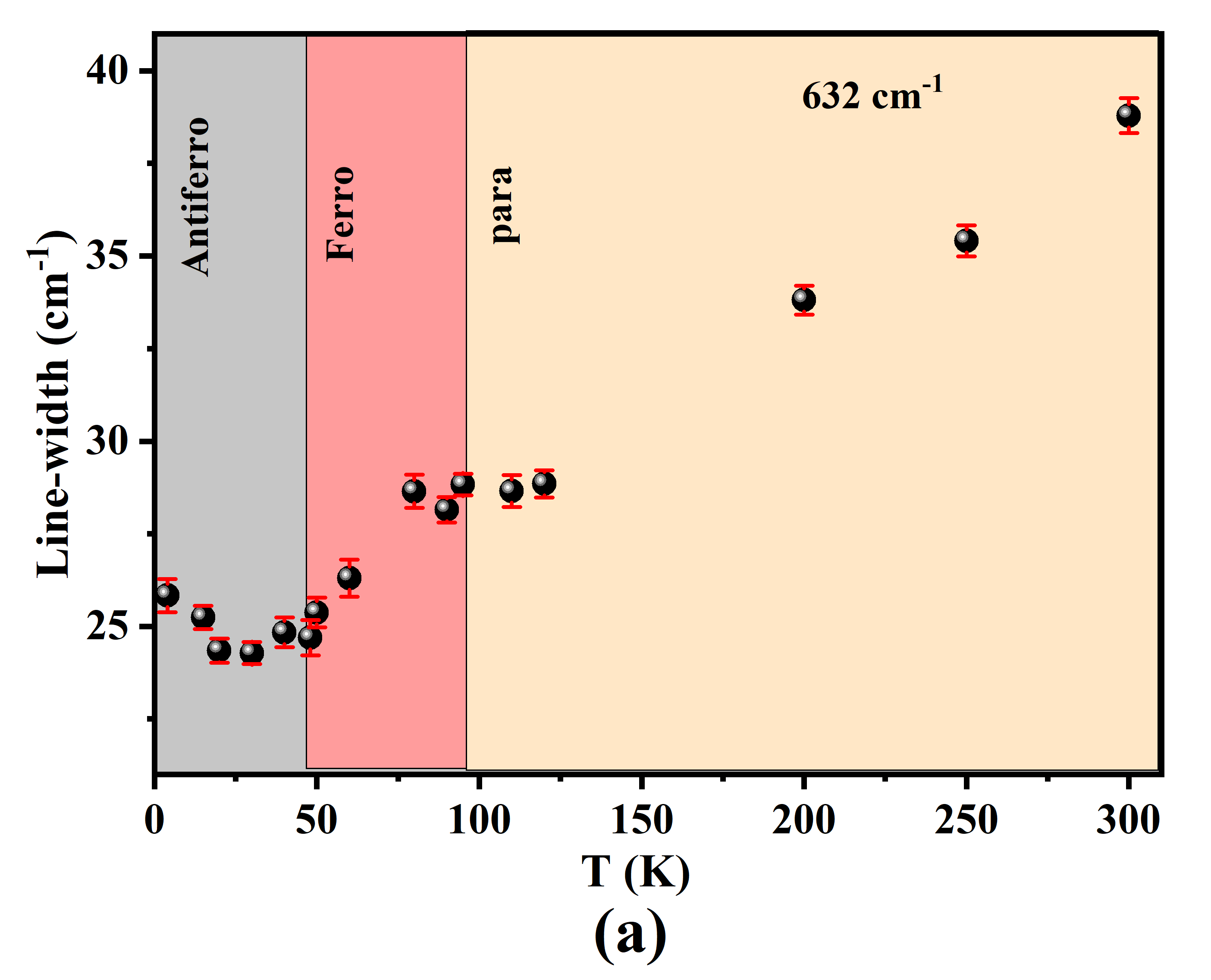}  
	\includegraphics[scale=.3]{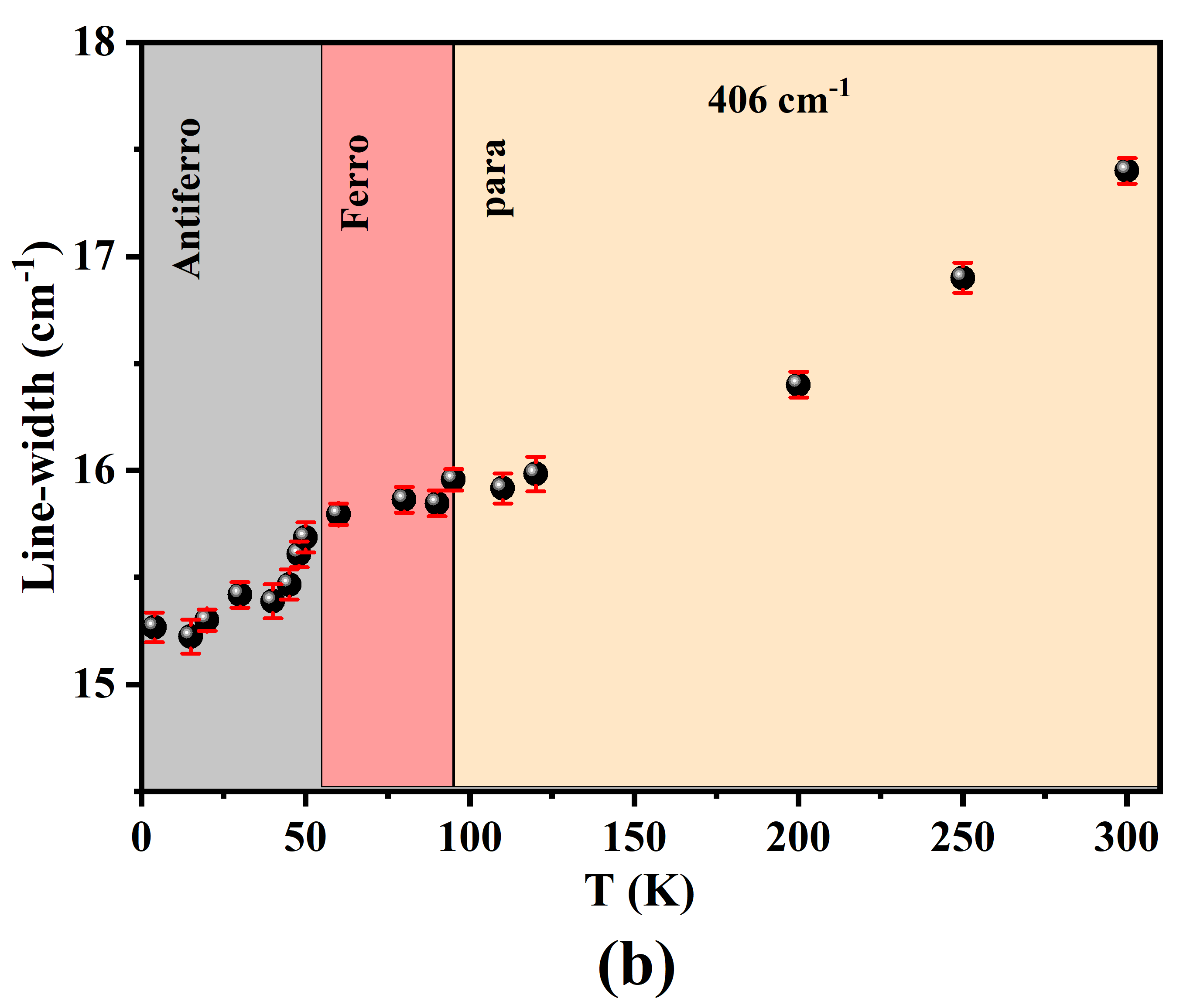}
	\caption{Temperature-dependence of Raman line width (FWHM) of (a) A$_g$ mode, and (b) B$_g$ mode. The error bars are the standard deviation of the peak positions as obtained from the fitting procedure.}
 \label{raman4}
\end{figure*}

Now, we investigate the role of magnetostriction on the observed phonon anomaly. In this regard, the study of FWHM or Raman linewidth is more effective than the position with temperature. Raman linewidth is associated with phonon life-time that is not affected by lattice volume changes, i.e., magnetostriction effect. Moreover, FWHM is affected by magnetic transition and reveals an anomaly in the FWHM near the transition. Therefore, FWHM is more effective for determining whether the anomaly arises from direct spin-phonon coupling or is induced by magnetostriction.  Fig.~\ref{raman4} (a-b) shows the temperature dependence of the linewidth of the showing anomaly by stretching and anti-stretching modes of YGCMO. Generally, a significant anharmonic effect has been observed at higher temperatures due to the rigorous atomic vibration of the atoms about their mean position. However, this anharmonic effect is reduced with a decrease in temperature due to a decrease in phonon-phonon interactions. As a result, the phonon lifetime increases or the FWHM decreases. We observed that both curves show anomaly upon entering into the magnetic region with the discontinuities at 95.5 K ($T_C$) and 47 K ($T_N$), respectively, suggesting the relative dependence of phonon lifetime with magnetic ordering. Therefore, strong interaction between phonon and magnetic degree of freedom leads to an increase in linewidth at lower temperatures i.e., a decrease in phonon lifetime. Usually, the linewidth is affected by spin-phonon coupling and electron-phonon coupling \cite{RD_21}. The total electronic density of states for YGCMO shows that it is half-metallic (refer to Fig.S5 (b) in SI). Therefore, the anomaly in linewidth with temperature is ascribed to spin-phonon coupling \cite{RD_21}. The careful observation of the above temperature-dependent frequency and linewidth plot reveals a new phonon renormalization arising at T$_N$ due to AFM ordering of Gd-Co/Mn spins, implying the presence of spin-phonon coupling. \\
\begin{figure*}[htp]
	\centering
	\includegraphics[scale=.3]{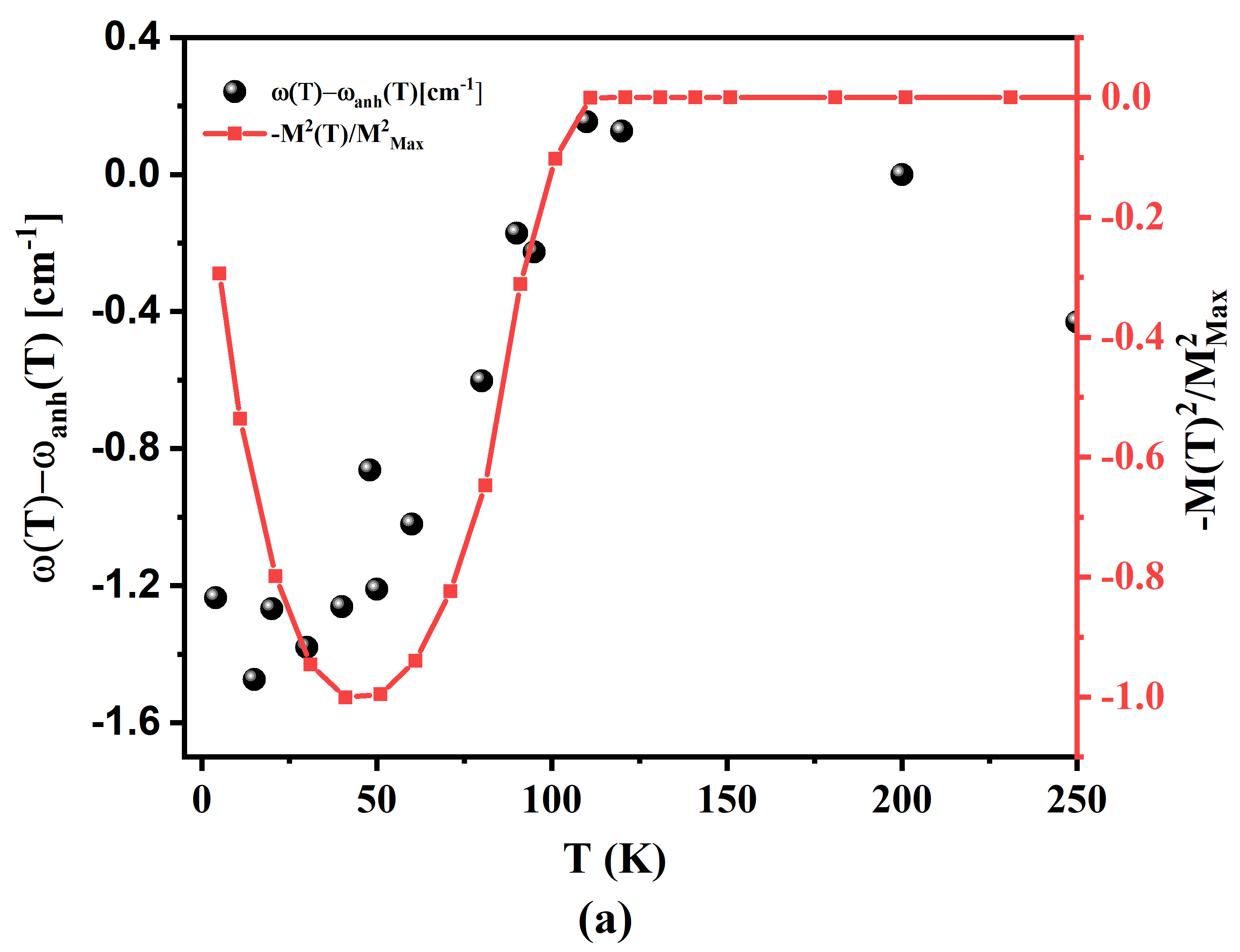}  
	\includegraphics[scale=.3]{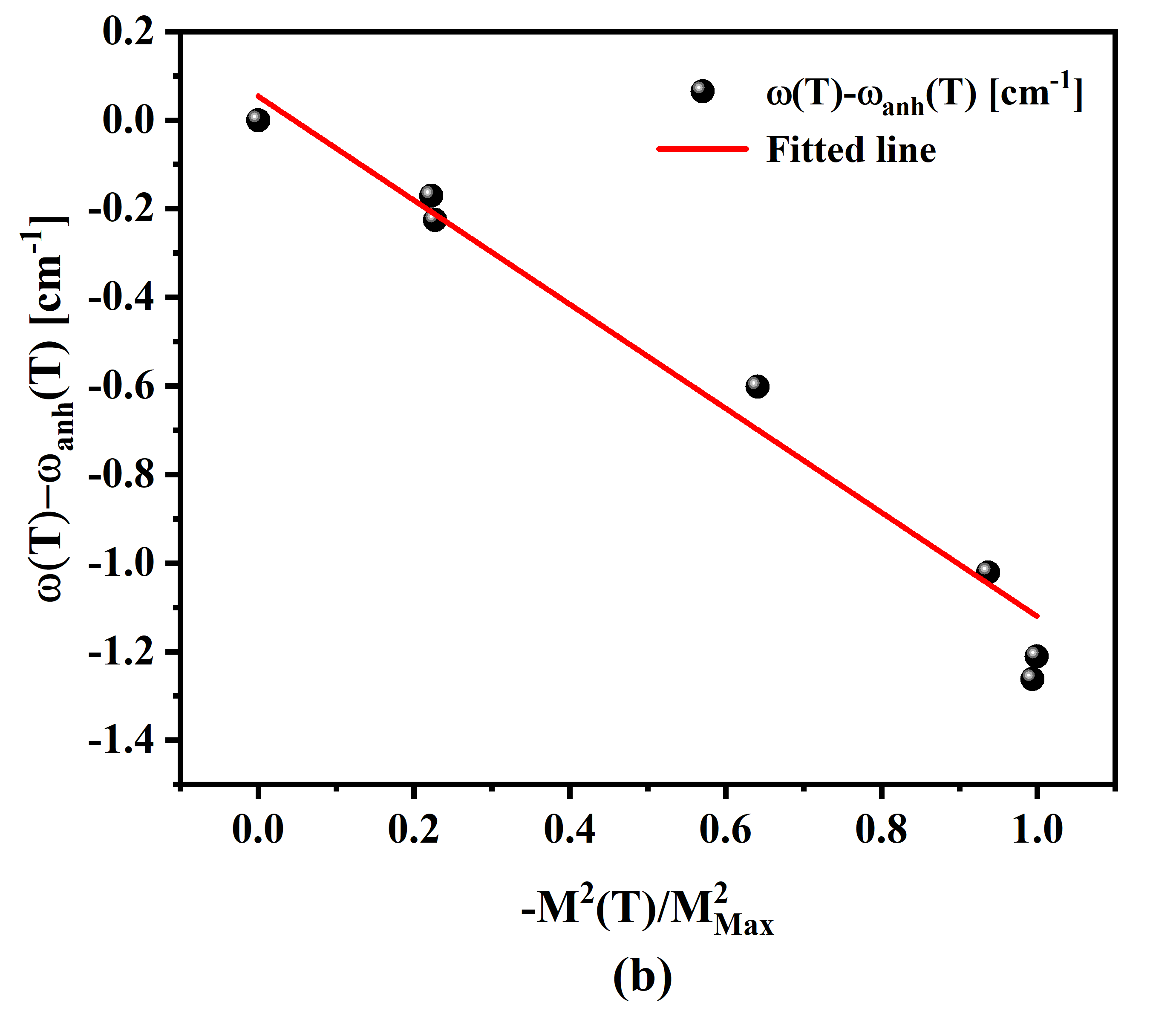}
	\caption{(a) A comparison of the deviation ($\Delta \omega (T)$) of A$_g$ stretching mode from the anharmonic model and the normalized magnetization $\frac{M^2(T)}{M^2_{max}}$. (b) the corresponding linear dependence of $\Delta \omega (T)$ on the squared relative magnetisation $\frac{M^2(T)}{M^2_{max}}$.  }
	\label{raman5}
\end{figure*}
\noindent To get detailed insight into the anomalous behavior of Raman modes, the renormalization of phonon frequency with normalized magnetic moment has been plotted in Fig.~\ref{raman5}(a).
Usually for magnetic oxide, the deviation of phonon mode from the anharmonic model can be explained by the spin-spin correlation function in magnetic oxides \cite{int_28}. 
\begin{equation}
    \Delta \omega_{SPC}(T)= \omega(T) - \omega_{anh}(T) \approx -\lambda \langle {\bf{S_i} \cdot S_j} \rangle
\end{equation}
where $\omega(T)$ is the renormalized phonon frequency due to SPC, $\omega_{anh}$ is the phonon frequency in the absence of SPC, $S_i$ and $S_j$ are nearest neighbor spins at $i^{th}$ and $j^{th}$ sites and $\langle {\bf{S_i} \cdot S_j} \rangle$ is the statistical average of the spin pair correlation function, respectively. $\lambda$ denotes the strength of spin-phonon coupling. $\Delta \omega$ is obtained by taking the difference between calculated and observed phonon frequencies. According to mean-field theory, the spin-spin correlation function can be connected to magnetization by the equation $\Delta \omega_{SPC} \approx$ $(M(T)/M_{max})^2$, where $M(T)$ and $M_{max}$ are the magnetization at T and maximum magnetization, respectively. Therefore, by considering the four nearest neighbors for each B site cation, $\Delta\Omega_{SPC}$ can be written as follows \cite{RD_17}: 
\begin{equation}
	\Delta \Omega_{SPC} \approx -\lambda \langle {\bf{S_i} \cdot S_j} \rangle \approx -4\lambda \frac{M^2(T)}{M^2_{max}}
\end{equation}
The spin-phonon coupling constant value $\lambda$ is estimated by using the above approach. Its value can be positive or negative depending upon phonon hardening or softening. Fig.~\ref{raman5}(a) depicts that there is a good overlap of both $\Delta \omega_{SPC}$ and normalized magnetization $\frac{M^2(T)}{M^2_{max}}$ near the transition temperature, beyond which a deviation in the magnitude of $\Delta \omega_{SPC}$ occurs. It signifies that the origin of anomalous softening is ascribed to spin-phonon coupling. The occurrence of this unconventional behavior below 95 K might be caused by competition between FM and AFM interactions. In this sample, the ferromagnetic (FM) phases coexist with antiferromagnetic (AFM) phases, where the predominant AFM interactions, such as Co$^{3+}$-O- Co$^{3+}$, Mn$^{3+}$-O-Mn$^{3+}$, Co$^{2+}$-O- Co$^{2+}$, and Mn$^{4+}$-O-Mn$^{4+}$ exchange pathways are produced due to ASD.  Additionally, Gd-4f and Co/Mn-3d interaction is present in the sample. Thus, the magnetic frustration emerges from the competition between FM and AFM interactions.
Therefore, different values of magnetic coupling $J_{ij}$ can contribute in different ways to the phonon renormalization induced by magnetic ordering. It makes the spin-phonon coupling more complex and causes deviations from the mean-field approximation. A similar kind of deviation is observed in La$_2$CoMnO$_6$ \cite{RD_16}, Gd$_2$CoMnO$_6$ \cite{RD_17}, and Lu$_2$CoMnO$_6$ \cite{RD_18}, where FM and AFM interactions coexist together.  \\
Spin-phonon coupling strength ($\lambda$) is determined by taking a quarter of the slope of the linear relationship of $\Delta \omega_{SPC}$ vs. $\frac{M^2}{M^2_{max}}$ \cite{int_16} shown in Fig.~\ref{raman5} (b). The value is found to be 0.29 cm$^{-1}$. The spin-phonon coupling (SPC) strength is notably weaker in bulk systems compared to thin films, with a reported SPC value of $\lambda$ = 0.51 for bulk Pr$_2$CoMnO$_6$ \cite{int_16}. Silva \textit{et al.} reported that ordered bulk LCMO does not show SPC, whereas disordered Gd(Co$_{1/2}$Mn$_{1/2}$)O$_3$ exhibits a significant SPC \cite{RD_22}. Therefore, our sample, containing 38\% ASD as stated above, shows SPC, and DFT calculations further reveal that SPC is enhanced in the presence of ASD. Thus, B-site cationic ordering in YGCMO perovskite profoundly influences not only its magnetic properties but also its phonon dynamics and spin-phonon coupling strength $\lambda$.\\

\begin{figure*}[htp]
	\centering
	\includegraphics[scale=.4]{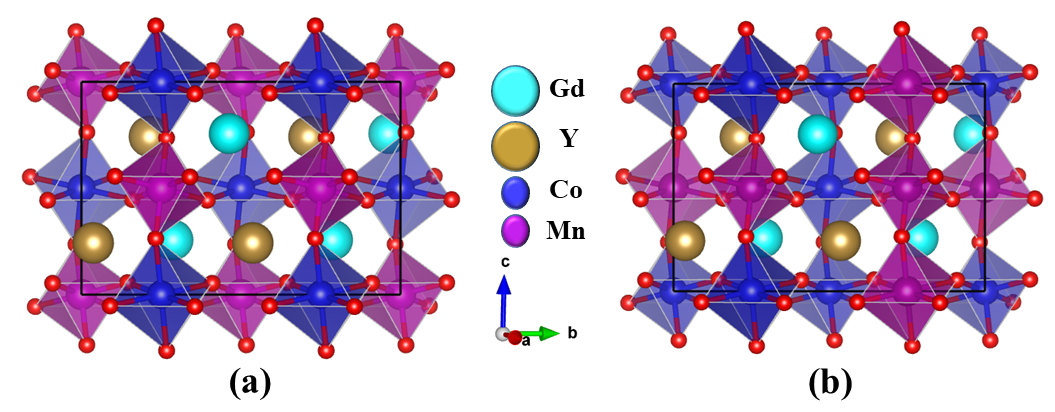}  
    \includegraphics[scale=.41]{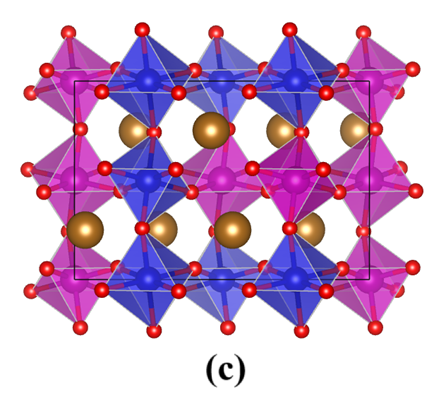}  
	\caption{Lowest energy crystal structure of 1$\times$2$\times1$ (a) ordered YGCMO, (b) disordered YGCMO, and (c) disordered YCMO. The blue and pink balls represent Co and Mn atoms, respectively, and sit at the centre of the octahedra. Here, Y and Gd atoms are represented by golden and cyan big balls, sitting in the cage formed by CoO$_6$ and MnO$_6$ octahedra.}
	\label{raman6}
\end{figure*}
The impact of Gd doping, Gd ordering, and the impact of ASD on Raman spectra and SPC in the YGCMO compound is further investigated by first-principles calculations. We have considered a supercell of size 1$\times$2$\times$1 containing 40 atoms of YCMO and YGCMO i.e, there are (8 Y, 4 Co, 4 Mn, 24 O) and (4 Y, 4 Gd, 4 Co, 4 Mn, and 24 O) atoms in each supercell. The structure was fully optimized, starting from the room temperature experimental data. Fig.~\ref{raman6}(a) shows the lowest energy ordered structure of YGCMO where Co and Mn octahedra are arranged alternately. To determine the influence of anti-site disorder on the nature of SPC, starting from the ordered 1$\times$2$\times$1 crystal structure, we have manually introduced some anti-site disorder. This disordered structure thereafter underwent a complete optimization (see Fig.~\ref{raman6} (b)). The total electronic density of states for the disordered structure (which contains Co-Co, Co-Mn, and Mn-Mn interactions) shows that the sample is half-metal with a band gap of 1.1 eV in down spin channel(refer to Figure S5 (b) of the SI). 
\begin{figure*}[h]
	\centering
	\includegraphics[scale=.4]{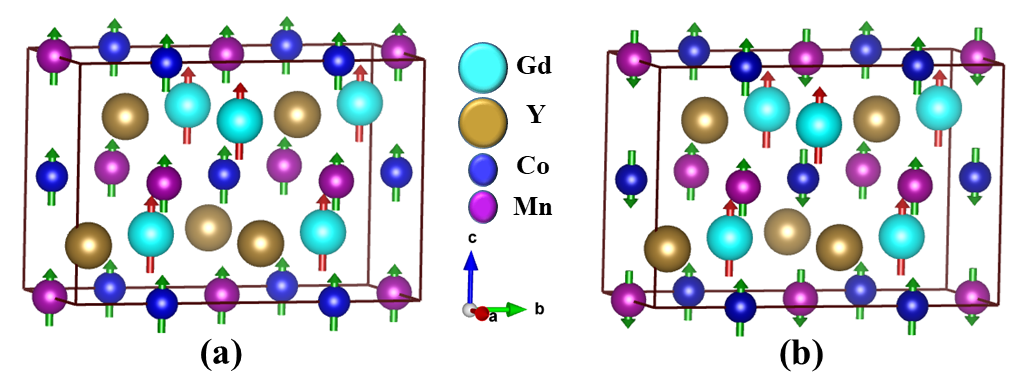}  
	\caption{Energetically preferred spin configurations of 1$\times$2$\times$1 ordered YGCMO (a) Ferromagnetic ordering; (b) C-type AFM ordering. Here, blue and pink small balls and golden and cyan big balls represent Co, Mn, Y, and Gd atoms, respectively. The red arrows denote the spins on Gd atoms, and green arrows represent the spins on Co and Mn atoms. }
	\label{raman7}
\end{figure*}
To determine the magnetic ground state, we have performed total energy calculations on the ferromagnetic, C-, G-, and A-type antiferromagnetic and ferrimagnetic configurations of ordered and disordered-YGCMO. These magnetic interactions were considered among the Co and Mn ions, whereas the Gd moments were made to orient along the z-direction in all the calculations. Our calculations indicate that the ferromagnetic state (see Fig.~\ref{raman7}(a)) has the lowest energy, followed by the C-type AFM magnetic configuration (see Fig.~\ref{raman7} (b)). The relative energies of the different magnetic orderings with respect to the ferromagnetic state have been tabulated in Table~\ref{Renergy}. 
\begin{table*}[h]
	\centering
	\caption{\bf{Total energy of different magnetic configurations of ordered YGCMO with respect to the ferromagnetic state.}}
	\begin{tabular}{|l | c |}
		\hline
		\hline
		Spin configurations& Relative energy values (eV/f.u.)\\
		\hline
		\hline
		FM  &	0.0\\
        C-type AFM & 0.214 \\
		G-type AFM & 0.259 \\
		A-type AFM  & 1.863 \\
		Ferrimagnetic & 0.252 \\
		\hline
		\hline
		\end{tabular}
  \label{Renergy}
\end{table*}
\begin{figure*}[h]
	\centering
	\includegraphics[scale=.34]{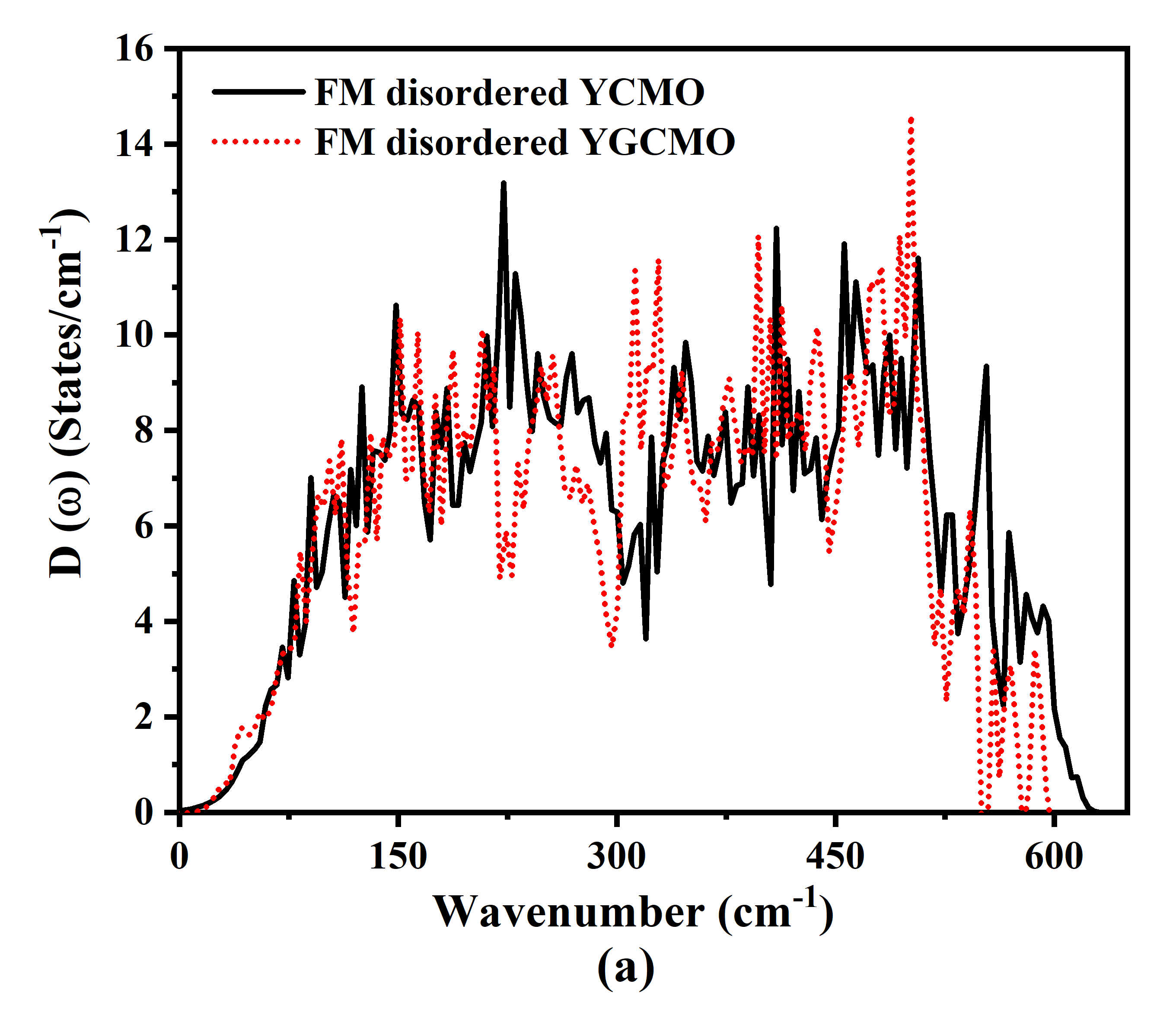}  
	\includegraphics[scale=.34]{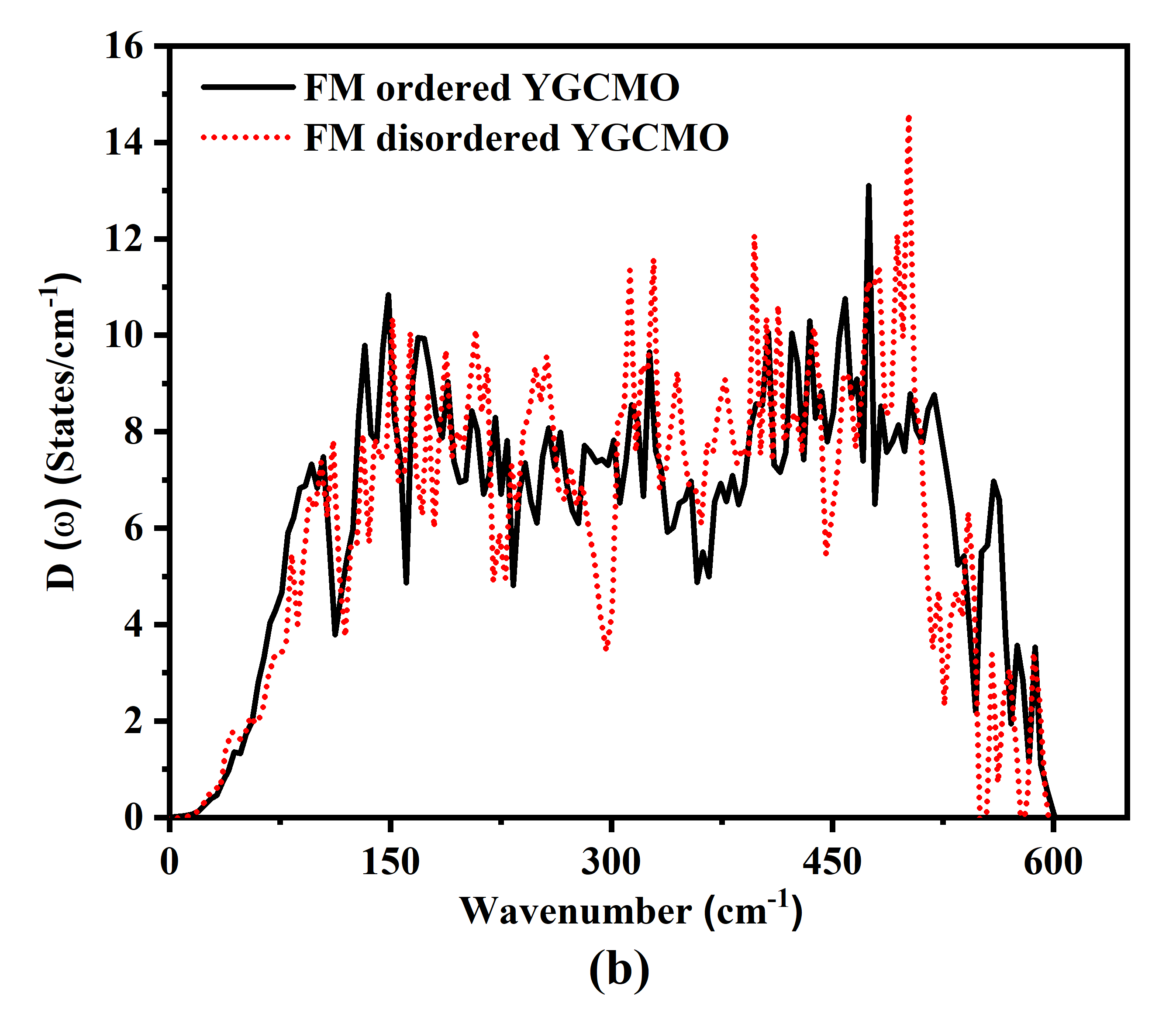}
	\caption{Comparative total phonon density of states of (a) disordered YCMO and YGCMO, (b) ordered and disordered YGCMO .}
	\label{raman08}
\end{figure*}
In order to understand the nature of the influence of Gd$^{3+}$ substitution at Y$^{3+}$ sites on the Raman spectra, 
 We calculated the total phonon density of states (TPDOS) for disordered YCMO and YGCMO. It is to be noted that analysis of the partial phonon density of states (PDOS) of YGCMO (as presented in Section S6(f) of the Supplementary Information) indicates that the translational and the librational modes related to the Gd$^{3+}$ ion appear in the lower wave number region (as Gd$^{3+}$ is heavier than Y$^{3+}$ by 76.88\% and wavenumber=$1/\sqrt{m}$ ($m$=mass)). A comparison of our calculated total phonon density of states for YCMO and YGCMO as presented in Fig.~\ref{raman08}(a) shows that the overall TPDOS gets shifted to lower frequencies with Gd doping. This happens because of change in bond length and bond angle with Gd doping (refer to Table~\ref{Gd-doping}). Thus, we conclude that A-site doping with a cation of a larger ionic radius shifts the phonon modes to lower frequencies.
\begin{table*}[h]
	\centering
	\caption{\bf{The average bond angles between transition metal elements and average bond length of Co-O and Mn-O obtained from DFT calculations are as follows}}
	\begin{tabular}{|l | c | c | c | c | c |}
		\hline
		\hline
   \multirow[c]{2}{*}{} & \multicolumn{3}{|c|} {{\bf Bond angle}} & \multicolumn{2}{|c|}{{\bf Bond length}}\\
      \cline{2-6}
		& Co-O-Mn ($^\circ$)& Co-O-Co ($^\circ$) & Mn-O-Mn ($^\circ$)& Co-O (\AA) & Mn-O (\AA)\\
	\hline
        \hline
		disordered YCMO   & 146.34 & 144.31 & 142.92 &  2.0241  & 1.9543  \\
        disordered YGCMO & 146.66&147.28&145.16 & 2.0426 & 1.9731\\
        \hline
        \hline
		\end{tabular}
  \label{Gd-doping}
\end{table*}
 With 50\% of the Y being replaced by Gd, one can very well appreciate that it will give rise to various different configurations depending upon how the Gd is arranged in the YCMO lattice. Hence, we have investigated the influence of Gd ordering on the Raman spectra of YGCMO. Apart from Gd-Y ordering at A site, we also encounter Co-Mn ordering at the B site of the double perovskite. Our calculations reveal that any alteration in Gd ordering modifies the total phonon density of states (TPDOS) or Raman spectra (refer to S8 with in the SI).
 
 We also calculated the PDOS of ordered and disorderd YGCMO to unveil the effect of antisite disorder on Raman spectrum. As illustrated in Fig.~\ref{raman08} (b), the phonon density of states (PDOS) for ordered and disordered YGCMO shows notable differences across the frequency spectrum. Antisite disorder at the B-site introduces Co-O-Co and Mn-O-Mn bonds, which increase lattice strain. This disorder modifies bond lengths and angles (refer to Table~\ref{bond_angle} and Table~\ref{bond_length}) (as wavenumber $= 1/\sqrt{l}$ ($l$=bond length)), thereby altering the PDOS. These structural changes directly influence the Raman spectrum.

To investigate whether there is a coupling between the spin and the lattice degrees of freedom, we have calculated the total phonon density of states for the two lowest energy magnetic configurations, namely, the ferromagnetic and the C-type antiferromagnetic state, in both its ordered and anti-site disordered state (as shown in Fig.~\ref{raman6}).
\begin{figure*}[h]
	\centering
	\includegraphics[scale=.34]{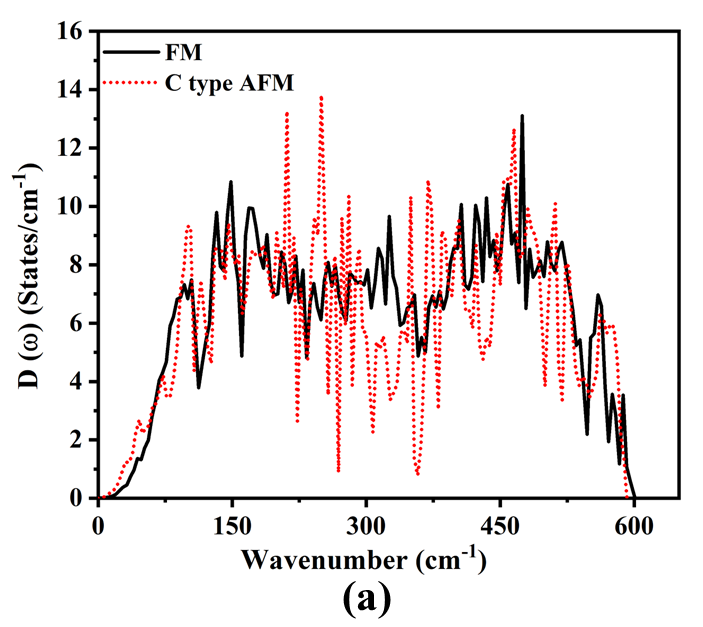}  
	\includegraphics[scale=.34]{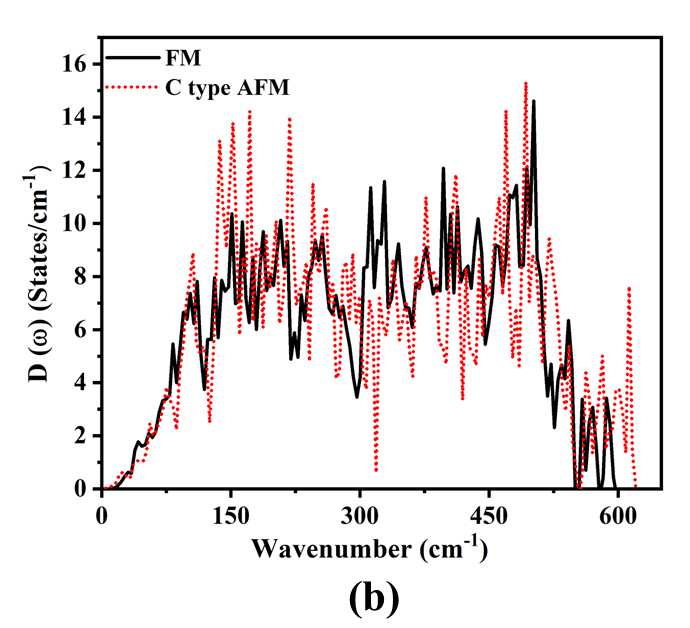}
	\caption{Comparative total phonon density of states of FM and C-type AFM for (a) ordered state, (b) disordered state.}
	\label{raman8}
\end{figure*}
\begin{table*}[htp]
	\centering
	\caption{\bf{The average bond angles between transition metal elements obtained from DFT calculations are as follows}}
	\begin{tabular}{|l | c | c | c |}
		\hline
		\hline
		& Co-O-Mn ($^\circ$)& Co-O-Co ($^\circ$) & Mn-O-Mn ($^\circ$)\\
		\hline
		\hline
		FM-ordered  & 144.85 & - & -\\
        C-type AFM-ordered & 145.31 & - & - \\
        \hline
        \hline
		FM-disordered & 146.66&147.28&145.16 \\
		C-type AFM disordered  & 144.78&145.09& 144.82 \\
		\hline
		\hline
		\end{tabular}
  \label{bond_angle}
\end{table*}
The total phonon density of states for these two configurations is shown in Fig.~\ref{raman8}. The absence of imaginary modes affirms the dynamical stability of both the magnetic configurations. It is evident from Fig.~\ref{raman8}(a) that with the change in magnetic ordering, there is an alteration in the phonon density of states, indicating that spin-phonon coupling is active in this system. The partial phonon density of states for FM and C-type magnetic configuration is presented in Figure S6 of the SI. We have also checked the influence of disorder at Co and Mn on the spin-phonon coupling, as presented in Fig.~\ref{raman8}(b).
\begin{table*}
	\centering
	\caption{\bf{The average bond length of Co-O and Mn-O obtained from DFT are as follows}}
	\begin{tabular}{|l | c | c | c | c |}
		\hline
		\hline
		& Co-O (\AA)& Std. deviation (\AA) & Mn-O (\AA)&Std. deviation (\AA)\\
		\hline
		\hline
		FM-ordered  & 2.0809 & 0.00 & 1.9444 &0.00\\
		C-type AFM-ordered &2.0926  & 0.0011 & 1.9400 & 0.0006\\
		\hline
		\hline
		FM-disordered & 2.0426 &0.05537 &1.9731  &0.02856\\
		C-type AFM disordered  &2.0447 &0.05997 &1.9567  &0.04715\\
		\hline
		\hline
	\end{tabular}
	\label{bond_length}
\end{table*}
Our calculations indicate that with the introduction of disorder in the system, the shift between prominent modes further increases approximately by 3\%, whereas it is found that the shift has been increased with an increase in Gd concentration. Specifically, for Gd$_2$CoMnO$_6$, the shift rises by 7\% as shown in S9 (b) of the SI, indicating an enhanced spin-phonon coupling (SPC). To investigate the cause behind the enhanced spin-phonon coupling in the case of disordered YGCMO, we performed a detailed structural analysis of the sample in the presence/absence of disorder. It was observed that in the ordered sample the Co-O-Mn bond angle was enhanced by 0.46$^{\circ}$ as the magnetic configuration was changed from FM to C-type. However, the same was found to be reduced by 1.88$^{\circ}$ for the disordered sample (refer to Table~\ref{bond_angle}). Analysis of the bond lengths of the CoO$_6$ and MnO$_6$ octahedra showed that in ordered YGCMO the average Co-O bond length is found to increase by 0.0117 \AA~ and the average Mn-O bond length was found to decrease by 0.0044 \AA~ (refer to Table~\ref{bond_length}) as the magnetic configuration was changed from FM to C-type AFM. On the contrary, for similar magnetic transformations, both the average Co-O and Mn-O bond lengths were found to get enhanced and reduced by 0.0021 \AA~ and 0.0164 \AA~, respectively (refer to Table~\ref{bond_length}). However, there is in general an increase in the distortion of the CoO$_6$ and MnO$_6$ octahedra with change in magnetic ordering, which is found to increase with the introduction of anti-site disorder. It has been observed previously \cite{bond_length} that the relation between the frequencies of the Raman modes ($\nu$) and the bond length ($l$) is $\nu = 1/\sqrt {l}$. Therefore, the modulation in bond length of a specific atom engaged with a particular vibration can lead to shifts in mode positions. This is also evident in our first-principles calculations, which show that in the presence of anti-site disorder, the magnetism induced distortion of the lattice is more, which eventually gets reflected in the phonon spectra.
\section{Conclusion}
In summary, temperature-dependent Raman study confirms that the sample holds its $P2_1/n$ symmetry throughout the investigated temperature range. The anomalous softening of A$_g$ stretching mode implies an emergence of SPC from the presence of both FM and AFM interactions. 
Again, the nature of phonon linewidth curves and the half-metallic nature of the material repudiate any effect of magnetostriction on the observed softening of A$_g$ modes. 
 On the other hand, shift in the A$_g$ mode position from the anharmonic approach exhibits an unconventional behavior with the square of the magnetization. This is because of competing FM and AFM magnetic ordering, which is driven by ASD. We have found from DFT calculations that the Gd substitution at A-site induces a shift of phonon modes toward the lower frequency region. Additionally, antisite disorder (ASD) and Gd ordering at the A-site significantly modify the Raman spectra. Our calculations also indicate the presence of SPC in both ordered and disordered YGCMO. The calculated value of the SPC strength is 0.29 cm$^{-1}$. 
 Notably, Gd doping at the A-site and ASD at the B-site significantly enhance SPC in the system.
 Microscopic analysis indicates an intricate relationship between magnetism, state of orderliness at the crystallographic sites, and the crystal structure of the YGCMO sample. Such materials are known to hold promise in the field of spintronics.
\section{Acknowledgments}
 AK acknowledges the University Grants Commission (UGC) and the Ministry of Education (MoE) for their financial support. AK, AT, and TKN also acknowledge the National Supercomputing Mission (NSM) for providing computing resources of PARAM Shakti at IIT Kharagpur, which is implemented by C-DAC and supported by the Ministry of Electronics and Information Technology (MeitY) and Department of Science and Technology (DST), Government of India. AS would like to thank the IIT Mandi for research facilities. DB acknowledges funding from the Prime Minister's Research Fellowship scheme (PMRF) and computing facilities of IISER Kolkata. SC acknowledges funding through the IISER Kolkata start-up grant and SERB-POWER Grant (Grant No.: SPG/2021/003842).


\end{document}